\documentclass[prb,aps,superscriptaddress,twocolumn]{revtex4}

\usepackage{graphicx}
\usepackage{dcolumn}
\usepackage{bm}
\usepackage{amssymb}

\begin{document}


\title{Model for frustrated spin-orbital chains: application to CaV$_2$O$_4$}

\author{Gia-Wei Chern}
\affiliation{Department of Physics, University of Wisconsin,
Madison, Wisconsin 53706, USA}
\author{Natalia Perkins}
\affiliation{Department of Physics, University of Wisconsin,
Madison, Wisconsin 53706, USA}

\date{\today}

\begin{abstract}
Motivated by recent interest in quasi-one-dimensional compound
CaV$_2$O$_4$, we study the ground states of a spin-orbital chain
characterized by an Ising-like orbital Hamiltonian and frustrated
interactions between $S=1$ spins. The on-site spin-orbit interaction
and Jahn-Teller effect compete with inter-site superexchange leading
to a rich phase diagram in which an antiferro-orbital phase is
separated from the orbital paramagnet by a continuous Ising
transition. Two distinct spin liquids depending on the underlying
orbital order are found in the limit of small spin-orbit coupling.
In the opposite limit, the zigzag chain behaves as a spin-2 chain
with Ising anisotropy. The implications for CaV$_2$O$_4$ are
discussed.
\end{abstract}

\maketitle

Orbital degrees of freedom have been shown to play an important role
in understanding the electronic and magnetic properties of
transition metal oxides. \cite{imada98} This is particularly so for
frustrated magnets with partially filled orbitals. A well-studied
case is vanadium spinels $A$V$_2$O$_4$ where the $A$-site sublattice
is occupied by divalent ions such as Zn$^{2+}$ or Mn$^{2+}$, and the
$B$-site magnetic V$^{3+}$ ions form a {\em pyrochlore} lattice.
\cite{lee04,garlea08} It is known that geometrical frustration of
classical spins on such a lattice precludes simple N\'eel ordering
and gives rise to a highly degenerate ground state. Orbital ordering
in these compounds partially relieves the frustration by creating
disparities in nearest-neighbor exchange constant, hence setting the
stage for magnetic ordering at lower temperatures.
\cite{tsunetsugu03,ot04}

\begin{figure}
\centering
\includegraphics[width=0.99\columnwidth]{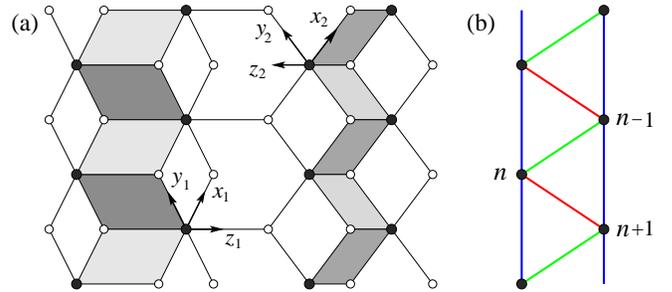}
\caption{\label{fig:v-chains} (Color online) (a) Two
crystallographically inequivalent vanadium chains in CaV$_2$O$_4$.
The V$^{3+}$ ions are arranged in zigzag chains of edge-sharing
VO$_6$ octahedra. For both vanadium sites, the VO$_6$ octahedra is
flattened at room temperatures. A local reference frame is defined
in such a way that $z$-axis is parallel to the tetragonal axis of
the crystal field.
The black and white circles denote the vanadium and oxygen ions,
respectively.
(b) A simplified view of the zigzag chain. The red,
green, and blue bonds are parallel to the local $[011]$, $[101]$ and
$[110]$ axes, respectively. Consequently, electron hopping on the
red, green, and blue bonds is possible only when $d_{yz}$, $d_{zx}$
and $d_{xy}$ orbitals are occupied, respectively.}
\end{figure}

In this paper, we investigate the physics of frustrated vanadium
chains in which the interplay of geometrical frustration,
spin-orbital couplings, Jahn-Teller effect, and enhanced quantum
fluctuations leads to a rich phase diagram. Our work is partly
motivated by an attempt to understand another vanadium compound
CaV$_2$O$_4$, which at room temperature crystalizes in the
orthorhombic calcium-ferrite structure (space group $Pnam$).
\cite{zong08,pieper08,niazi08,hastings67,kikuchi01} Contrary to its
spinel cousins, V$^{3+}$ ions in CaV$_2$O$_4$ are arranged in zigzag
chains of edge-sharing VO$_6$ octahedra (Fig.~\ref{fig:v-chains}).
Antiferromagnetic interaction on zigzag chains consisting of
triangular loops is subject to geometrical frustration as well. The
rather weak and frustrated inter-chain couplings make the vanadium
chains quasi-1D systems susceptible to quantum fluctuations.
Couplings of vanadium orbitals to spins and phonons add yet another
dimension to the intriguing physics of zigzag chains.

In CaV$_2$O$_4$, the zigzag geometry results in a spin-1 chain with
comparable nearest and next nearest-neighbor interactions. Combined
with the observation that $3d^2$ configuration of V$^{3+}$ ions tend
to have an easy-plane anisotropy, \cite{kikuchi01} the quasi-1D
compound CaV$_2$O$_4$ has been a favorable candidate for the
long-sought chiral spin liquid where a long-range chiral order
coexists with algebraically decaying spin correlations.
\cite{kikuchi01,hikihara00} However, as first pointed out by Pieper
{\em et al.}, \cite{pieper08} orbital degeneracy in this compound
drastically changes the above picture. Since the $t_{2g}$ levels are
split into a singlet and a doublet due to a flattened VO$_6$
octahedron, with the low-energy singlet always occupied, a double
degeneracy remains for the other electron. The orbital degrees of
freedom in V$^{3+}$ ion is thus described by an Ising-like variable.

Recent experimental studies on CaV$_2$O$_4$ seem to rule out a
possible chiral liquid phase as well. A structural transition at
$T_s \approx 141$ K reduces the crystal symmetry to monoclinic
$P2_1/n$. \cite{niazi08,pieper08} It was suggested in Ref.
\onlinecite{pieper08} that the simultaneous orbital ordering
relieves the magnetic frustration of zigzag chains. As the
inter-chain frustration is also lifted by the lattice distortion, a
three-dimensional N\'eel order sets in at $T_N \approx 71$ K. The
collinear spins are parallel to the crystal $b$-axis as evidenced by
both nuclear magnetic resonance and neutron diffraction
measurements. \cite{hastings67,zong08,pieper08}

To make progress toward an understanding of the ground-state
structure and the nature of phase transitions in CaV$_2$O$_4$, here
we study the zero-temperature phase diagram of its building blocks,
i.e. zigzag chains with $S=1$ spins and Ising orbital variables. We
propose a theoretical model which includes the superexchange (SE)
interaction, relativistic spin-orbit (SO) coupling, and Jahn-Teller
(JT) effect. We find that antiferro-orbital order favored by an
Ising-like orbital exchange is destroyed in the presence of large
on-site spin-orbit or Jahn-Teller coupling. Depending on the
underlying orbital configuration, magnetic properties of the zigzag
chain is equivalent either to those of two weakly coupled $S=1$
chains, or of an unfrustrated spin-1 ladder. In the limit of large
spin-orbit coupling, the zigzag chain can be viewed as a spin-2
chain with anisotropic interaction. Finally, we discuss implications
for CaV$_2$O$_4$.

{\em Model Hamiltonian}. We first define a local reference frame for
the two crystallographically inequivalent vanadium chains (referred
to as type-I and II chains here) in CaV$_2$O$_4$ such that the
$z$-axis is parallel to the tetragonal direction of flattened VO$_6$
octahedron [Fig. \ref{fig:v-chains}(a)]. Nearest-neighbor bonds
along the chain are parallel to the local $[011]$ and $[101]$
directions alternatively. In the following, we employ a convention
in which even-numbered bonds are along the $[011]$ axis.

Geometrical frustration of the zigzag chain comes from the fact that
second-nearest-neighbor bonds parallel to local $[110]$ axis have a
length close to that of nearest-neighbor bonds. In fact, the
second-nearest-neighbor interaction is the dominant one as $d_{xy}$
orbital is occupied at every site due to the flattened VO$_6$
octahedra. The remaining orbital degeneracy is described by a
pseudospin-$\frac{1}{2}$ with $\sigma^z = \pm 1$ corresponding to
the $|yz\rangle$ and $i|zx\rangle$ states, respectively.

Having introduced the basic notations, we now discuss a minimal
model for the  spin-orbital chain. Since orbital interaction with a
90$^\circ$ angle between vanadium-oxygen bonds is governed by direct
$dd\sigma$ exchange of $t_{2g}$ orbitals, the corresponding spin
interaction on a bond depends on whether the relevant orbital is
occupied. \cite{dimatteo05} Essentially, orbitals participate in the
superexchange only via orbital projectors $P_{yz/zx}$. Noting that
$P_{yz/zx} = (1\pm \sigma^z)/2$, we first define the
antiferro-orbital and ferro-orbital bond operators
\[
    \mathcal{O}_{n,n+1} = \frac{1}{2}(1-\sigma^z_n\sigma^z_{n+1}),
    \quad
    \bar\mathcal{O}_{n,n+1} = \frac{1}{4}
    (1\pm\sigma^z_n)(1\pm\sigma^z_{n+1})
\]
with $\pm$ signs for even and odd numbered bonds, respectively. The
model Hamiltonian is divided into three parts: $H = H_0 + H_{\perp}
+ H_{\mbox{\scriptsize on-site}}$. The first $H_0$ term represents
decoupled spin and orbital systems
\begin{eqnarray}
    \label{eq:H0}
    H_0 = J_2 \sum_{n} \mathbf S_n\cdot\mathbf S_{n+2} - \sum_n
    (K\mathcal{O}_{n,n+1} +  K'\bar\mathcal{O}_{n,n+1}).
\end{eqnarray}
Since $J_2$ couples every second-nearest neighbors, the spin part
can be viewed as two decoupled $S=1$ Haldane chains, corresponding
to the two blue lines in Fig. \ref{fig:v-chains}(b). The $K$ and
$K'$ terms denote the energy gain of an antiferro-orbital and a
ferro-orbital bond, respectively. In general, we have $K > K'$ due
to a finite on-site Hund's coupling $J_H$, hence an
antiferro-orbital order in the ground state of $H_0$.

The $H_{\perp}$ term introduces interactions between the two spin-1
chains:
\begin{eqnarray}
    \label{eq:Hperp}
    H_{\perp} = \sum_n \bigl(J_{1}\bar\mathcal{O}_{n,n+1}
    - J'_{1} \mathcal{O}_{n,n+1}\bigr)\,
    \mathbf S_n\cdot\mathbf S_{n+1}.
\end{eqnarray}
As discussed above interaction between spins depends on orbital
occupations: the antiferromagnetic coupling $J_{1}$ is nonzero only
when $d_{yz}$ ($d_{zx}$) orbital is occupied at both sites of an
even (odd) bond, whereas the strength of ferromagnetic $J'_{1}$ term
depends on the expectation value of the antiferro-orbital bond
operator $\mathcal{O}_{n,n+1}$.

Explicit expressions relating the exchange constants to microscopic
parameters are obtained from the SE Hamiltonian of vanadium spinels,
\cite{dimatteo05} where neighboring VO$_6$ octahedra share the same
edge as in zigzag chains considered here. Assuming an exact
octahedral site-symmetry, we find $K= (1+2\eta)\frac{t^2}{U}$, $J_2
= J_1 = K' = (1- \eta) \frac{t^2}{U}$, and $J'_1 = \eta
\frac{t^2}{U}$ to lowest order in Hund's parameter $\eta \equiv
J_H/U$. Here $t$ is the  hopping integral, and $U$ is the on-site
Coulomb repulsion. The parameters of the model can be estimated from
known values of the same parameters in other vanadium compounds.
Measurements on cubic vanadates yield $J_H \simeq 0.68$ eV, $U\simeq
6$~eV, and $t\simeq -0.35$~eV, \cite{mizokawa96,takubo06} which
gives $\eta \simeq 0.11$ and $t^2/U\simeq 20.4$~meV. The estimate of
$\lambda$ is less certain, we find $\lambda\simeq 13-25$~meV.
\cite{abrag70,ot04,tanaka02,mizokawa96} In the following, we shall
measure energy in units of $t^2/U$.

The remaining single-ion interactions are included in the
Hamiltonian
\begin{eqnarray}
    \label{eq:Honsite}
    H_{\mbox{\scriptsize on-site}}
    &=& -\lambda \sum_n \sigma^x_n S^z_n
    - \delta \sum_n \sigma^z_n.
\end{eqnarray}
The first term originates  from relativistic SO coupling
$\lambda(\mathbf L \cdot \mathbf S)$. Since $d_{xy}$ orbital is
always occupied, the $x$ and $y$ components of the orbital angular
momentum are quenched; the remaining $L^z = -\sigma^x$ in the
pseudospin representation. A similar situation has been studied in
cubic vanadates. \cite{horsch03} The effect of the monoclinic
structural transition at $T_s$ is modeled by the second term with
$2\delta$ denoting the level splitting due to  the induced
orthorhombic distortion of VO$_6$ octahedra. Note that the
orthorhombic distortion is different on type-I and II chains.
\cite{pieper08,niazi08}

{\em Orbital orders}. We first consider a simpler case of the model
Hamiltonian by assuming the presence of a collinear N\'eel order on
the zigzag chain.  This is a plausible assumption as the SO term
$\lambda \sigma^x_n S^z_n$ breaks the spin SU(2) symmetry and, as
will be discussed later, closes the energy gap of longitudinal
magnons at large enough $\lambda$, hence signaling a transition to
the N\'eel state with $S^z_n$ parallel to $\pm \hat \mathbf z$. Even
with this simplification, the competition between inter-site
exchange and various on-site interactions still poses a rather
nontrivial problem. This study also sheds light on orbital orders in
the ground state of CaV$_2$O$_4$, where spins develop a
three-dimensional collinear antiferromagnetic order at $T_N \approx
71\,\mbox{K}$.

\begin{figure}
\centering
\includegraphics[width=0.99\columnwidth]{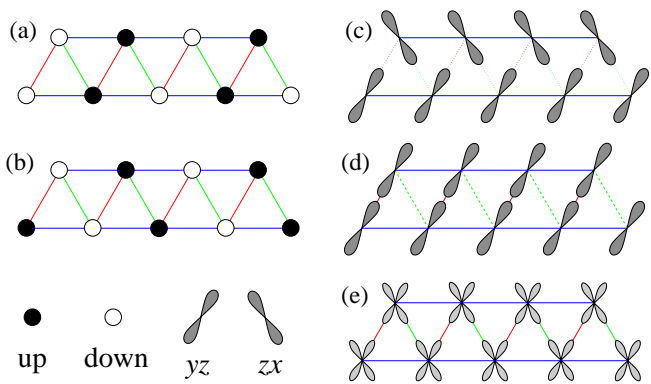}
\caption{\label{fig:orbital-orders} (Color online) Spin and orbital
orders on a zigzag vanadium chain. N\'eel orders (a) and (b) are
related to each other by lattice translations. There are a total of
four degenerate N\'eel states; the other two are related to states
(a) and (b) by time reversal. (c) antiferro-orbital order consisting
of staggered $d_{yz}$ and $d_{zx}$ orbitals. (d) and (e) correspond
to ferro-orbital orders with real orbital $d_{yz}$ and complex
orbitals $(d_{yz}\pm i d_{zx})/\sqrt{2}$, respectively.}
\end{figure}

Due to the strong second nearest-neighbor exchange $J_2$, collinear
orders on a zigzag chain consisting of repeated $++--$ spins have a
quadrupoled unit cell. There are a total of four degenerate N\'eel
states related to each other by lattice translations and time
reversal [Figs.~\ref{fig:orbital-orders}(a) and (b)]. After applying
a $\pi$-rotation about $z$-axis to pseudospins at $S_n^z<0$ sites,
we obtain an effective orbital Hamiltonian:
\begin{eqnarray}
    \label{eq:ising}
    H_{\rm orb} = \sum_n [\mathcal{J} + (-1)^n
    \mathcal{K} ]\sigma^z_n\sigma^z_{n+1}
    - \sum_n (h_z \sigma^z_n + h_x \sigma^x_n).\,\,
\end{eqnarray}
This model is equivalent to an Ising chain with alternating
nearest-neighbor couplings in a skewed magnetic field. The effective
exchange constants are $\mathcal{J} = (2K - K')/4$ and
$\mathcal{K}=\nu^2 (2J'_{1} + J_1)/4$. Here we have assumed $\langle
\mathbf S_n\cdot\mathbf S_{n+1}\rangle = \mp \nu^2$ on even and odd
bonds, respectively. The longitudinal and transverse fields are
given by $h_z = \delta + \nu^2 J_{1}/2$ and $h_x = \lambda$,
respectively. The parameter $\nu < 1$ characterizes the magnitude of
the N\'eel order. Its value can only be determined with a proper
treatment of the SO coupling term. For simplicity, we set $\nu = 1$
in the following calculation. Hamiltonian (\ref{eq:ising}) without
the staggered exchange $\mathcal{K}$ is one of the simplest models
exhibiting nontrivial quantum critical point: \cite{ovchinnikov03}
numerical calculation shows an order-disorder transition belonging
to 2D Ising universality class.  The staggered exchange
$\mathcal{K}$ involves higher-order spatial derivatives in the
continuum limit and thus represents an irrelevant perturbation in
the RG sense.

\begin{figure}
\centering
\includegraphics[width=0.97\columnwidth]{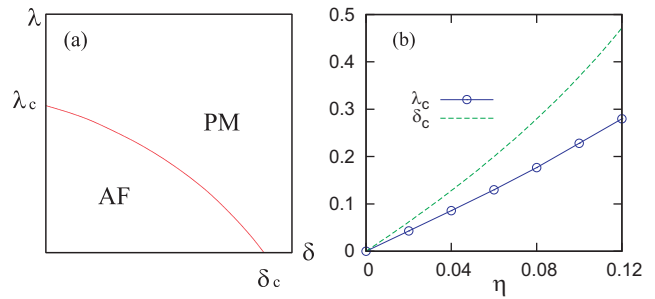}
\caption{\label{fig:phase-orbital} (a) Schematic phase diagram of
orbitals in the presence of a collinear N\'eel order on the zigzag
chain, effectively described by Hamiltonian (\ref{eq:ising}). AF and
PM refer to antiferro-orbitally  ordered and orbital paramagnetic
phases, respectively. (b) Critical boundaries $\lambda_c$ and
$\delta_c$ measured in units of $t^2/U$ versus Hund's parameter
$\eta$. The critical distortion is determined analytically from
$\delta_c = (2K-K'-J_1)/2$, whereas $\lambda_c$ is obtained from
DMRG calculation. The various effective parameters in Eq.
(\ref{eq:ising}) are expressible in terms of model parameters $J_1$,
$J'_1$, $J_2$, $K$, and $K'$ of the original SE Hamiltonian, whose
relations to $\eta$ and $t^2/U$ can be found in the text.}
\end{figure}

\begin{figure}
\centering
\includegraphics[width=0.8\columnwidth]{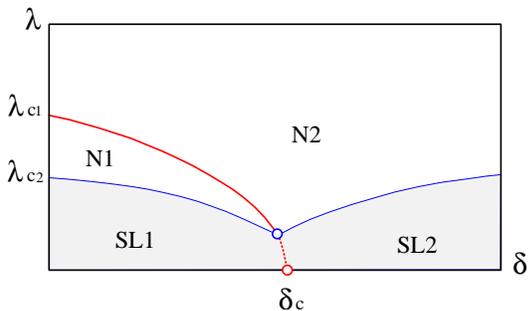}
\caption{\label{fig:phase} (Color online) Schematic phase diagram of
the zigzag vanadium chain. The various phases are characterized by
the magnetic properties.  SL and N represent spin liquid phase and
N\'eel order, respectively. The SL1 and N1 phases are accompanied by
an antiferro-orbital order, while in the N2 and SL2 phases the
orbitals behave as an orbital paramagnet. The first-order transition
along the dashed line is an extension of the multicritical point
$(\delta_c,0)$ in the phase diagram Fig. \ref{fig:phase-orbital}.}
\end{figure}

A schematic phase diagram of model (\ref{eq:ising}) is shown in
Fig.~\ref{fig:phase-orbital}(a), where an antiferro-orbital phase is
separated from the orbital paramagnet by an Ising transition line.
Along the $\delta$-axis ($\lambda = 0$), the antiferro-orbital phase
coexists with ferro-orbitally ordered phase, shown in
Figs.~\ref{fig:orbital-orders}(c) and (d) respectively, at the
multicritical point $\delta_c = (2K-K'-J_{1})/2$. On the other hand,
in the large $\lambda$ limit, the pseudospins are polarized by the
transverse field $h_x$ such that spins $S^z=\pm 1$ are accompanied
by complex orbitals $\frac{1}{\sqrt{2}}(d_{yz}\pm i d_{zx})$,
respectively, resulting in a uniform orbital occupation
$n_{yz}=n_{zx} = 1/2$ at all sites
[Fig.~\ref{fig:orbital-orders}(e)]. Using the infinite-system DMRG
method with periodic boundary condition, \cite{white93} we obtain a
critical $\lambda_c$ at $\delta=0$, taking into account the effect
of staggered $\mathcal{K}$. The Ising transition line is bounded by
the critical points $(\delta_c,0)$ and $(0,\lambda_c)$.
Fig.~\ref{fig:phase-orbital}(b) shows $\delta_c$ and $\lambda_c$ as
a function of parameter $\eta = J_H/U$. As expected, the
antiferro-orbital phase shrinks with decreasing Hund's coupling
$J_H$. In a full quantum treatment of the zigzag chain, we expect a
similar critical line characterized by massless orbital excitations
(Fig. \ref{fig:phase}).

{\em Spin liquid phases}. We now discuss the original Hamiltonian
(\ref{eq:H0})--(\ref{eq:Honsite}) in small $\lambda$ limit without
assuming the existence of a magnetic order. It is important to note
that the spin part of the Hamiltonian in this limit preserves a
continuous SU(2) symmetry, which can not be broken in one dimension.
One thus expects stable spin-liquid phases whose properties depend
critically on orbital configurations. Furthermore, the absence of
$\sigma^x$ term at $\lambda = 0$ indicates that the orbital part is
described by a classical Ising-like Hamiltonian. Consequently, the
search of ground states reduces to first enumerate over all possible
Ising configurations $\{\sigma^z_n\}$, and then compare their
energies taking into account contribution from spins. As
antiferro-orbital and ferro-orbital orders are favored by SE and JT
interactions, respectively, it is natural to consider these two
configurations first.

In the case of antiferro-orbital order
[Fig.~\ref{fig:orbital-orders}(c)] where bond operators $\langle
\bar\mathcal{O}_{n,n+1}\rangle = 0$ and $\langle
\mathcal{O}_{n,n+1}\rangle = 1$, the zigzag chain behaves as two
spin-1 chains weakly coupled by a frustrated ferromagnetic $J'_{1}$.
The corresponding spin liquid phase (SL1 phase in Fig.
\ref{fig:phase}) has an energy gap at an incommensurate wavevector.
\cite{allen00} On the other hand, the frustrated $J'_{1}$ coupling
is quenched by the ferro-orbital order shown in
Fig.~\ref{fig:orbital-orders}(d). Depending on the sign of
ferro-Ising order $\langle \sigma^z_n \rangle = \pm1$, the $J_{1}$
term is nonzero only on even or odd bonds, but not both. The spin
Hamiltonian is equivalent to that of a spin-1 ladder with rung
coupling $J_{1}$. The magnetic ground state is again disordered (SL2
phase in Fig. \ref{fig:phase}). \cite{senechal95,todo01} The
magnetic energy of spin-1 ladder with arbitrary rung coupling has
been calculated using quantum Monte Carlo method in Ref.
\onlinecite{senechal95}. Comparing the energy of the two phases,
including both spin and orbital contributions, yields a boundary
$\delta_c$ surprisingly close to the one obtained assuming a
preexisting N\'eel order.

Although the SU(2) symmetry is broken in the presence of SO
coupling, both spin liquids persist up to finite $\lambda$. For SL2
phase stable at $\delta > \delta_c$, SO coupling provides an
easy-axis anisotropy $D S_z^2$ with $D\simeq-\lambda^2/2\delta$. The
spin-1 ladder undergoes an Ising transition to a N\'eel phase (N2
phase in Fig.~\ref{fig:phase}) with increasing $\lambda$. At large
$\delta$, the condition $D=D_c$ gives rise to a critical $\lambda_c
\propto \sqrt{\delta}$. On the other hand, N\'eel transition at
small $\delta$ can be understood by considering the $\lambda$
dependence of the spin gap in SL1 phase. The singlet ground state of
weakly coupled Haldane chains has triply degenerate magnon
excitations with dispersion $\omega_k = \sqrt{\Delta_0^2 +
v^2(k-k_0)^2}$. In the limit of vanishing $J'_{1}$, the system
reduces to two decoupled Haldane chains with $k_0 = \pi$, $\Delta_0
= 0.410 J_2$ (Haldane gap for spin-1), and $v=2.49 J_2$.
\cite{affleck92} A simple second-order perturbation calculation adds
a correction to the spin gap
\begin{eqnarray}
    \Delta = \Delta_0 - \frac{\lambda^2 |\langle 0|S^z_n| k_0\rangle|^2}
    {\Delta_{\sigma} -\Delta_0},
\end{eqnarray}
where $\Delta_{\sigma} \approx 2(\delta_c-\delta)$ is the energy of
a domain-wall pair (a flipped pseudospin in the antiferro-orbital
state), $|0\rangle$ is the singlet ground state, and $|k\rangle$
denotes one-magnon excitation with wavevector $k$. The matrix
element $\langle 0|S^z_n|k\rangle = \sqrt{Z}e^{ikn}$, with $Z\approx
1$. \cite{affleck92} For $\Delta_{\sigma} > \Delta_0$, the spin gap
decreases with increasing $\lambda$ and eventually reaches  zero at
$\lambda_{c2} \propto \sqrt{2(\delta_c - \delta) - \Delta_0}$,
signaling a transition into magnetically ordered phase (N1 phase).

The two N\'eel phases in Fig.~\ref{fig:phase} are distinguished by
the underlying orbital orders, the phase boundary $\lambda_{c1}$
separating N1 from N2 phases is thus analogous to the Ising
transition line of the orbital only model
[Fig.~\ref{fig:phase-orbital}(b)]. Since the upper critical
$\lambda_{c1} \propto J_H$, at small Hund's coupling the zigzag
chain could bypass the N1 phase and enter the orbital paramagnetic
phase simultaneously with a magnetic ordering. A conjectured phase
diagram of the zigzag chain is shown in Fig. \ref{fig:phase}.

{\em Anisotropic $J=2$ spin chain}. In the limit of large SO
coupling, the appropriate degrees of freedom are effective spins of
length $J=2$, where $\mathbf J = \mathbf L' +\mathbf S$ and $L'=1$
is the angular momentum of the unoccupied $t_{2g}$ hole. The system
thus behaves as a spin-2 chain with anisotropic exchange
interaction. \cite{ot04} Furthermore, JT coupling adds an anisotropy
$D J_z^2 + E (J_x^2 - J_y^2)$, where $D<0$ and $E \propto \delta$
are proportional to the tetragonal and orthorhombic distortions of
VO$_6$ octahedron, respectively. Assuming a dominating easy-axis
anisotropy $|D| \gg E$, the N2 phase can also be viewed as N\'eel
ordering of the effective spins $J_z = \pm 2$ in such a way that
spins of a given direction $S_z=\pm 1$ is coupled to orbital angular
momenta $L'_z = \pm 1$, respectively. The corresponding
ferro-orbital order with complex orbitals $d_{yz}\pm i d_{zx}$ is
shown in Fig. \ref{fig:orbital-orders}(e). The V$^{3+}$ ions have a
reduced magnetic moment $\mu = (2S-L')\mu_B =1\mu_B$.

{\em Discussion}. We have presented and analyzed a minimal model of
frustrated vanadium chains, pertinent to quansi-1D compound
CaV$_2$O$_4$. A conjectured phase diagram (Fig.~\ref{fig:phase}) is
obtained based on analytical arguments and numerical calculations.
The observed $P2_1/n$ crystal symmetry of CaV$_2$O$_4$ at low
temperatures indicates that only {\em two} inequivalent vanadium
sites exist as in the high temperature phase. The absence of doubled
unit cell resulting from the antiferro-orbital order thus implies
that both vanadium chains are in the orbital paramagnetic phase.
This is a plausible conclusion noting that a rather small
$\lambda_{c1} \approx 0.22\,(t^2/U) \approx 4.5$ meV is estimated
from Fig.~\ref{fig:phase-orbital} assuming $\eta \approx 0.11$.
However, the monoclinic distortion below $T_s$ places the two types
of vanadium chain at rather distinct regions of the phase diagram.

The type-I chain acquires a small $\delta$ in addition to the
dominating tetragonal crystal field and behaves as a spin-2 chain
subject to an easy-axis anisotropy. This Ising anisotropy is
important to the stabilization of three-dimensional N\'eel order in
CaV$_2$O$_4$, as the collinear spins are found to point along the
easy axis of type-I chains. \cite{pieper08} The measured V moment
$1.06\mu_B$ is also consistent with the picture of an anisotropic
spin-2 chain. \cite{zong08} On the other hand, a strong orthorhombic
distortion $\delta$ at type-II ions completely removes the orbital
degeneracy and makes the vanadium chains effectively spin-1 ladders.
In fact, the well separated $t_{2g}$ levels at type-II ions lead to
a possible easy-plane spin anisotropy. \cite{pieper08} Consequently,
induced collinear order on type-II chains follows the spin direction
of type-I vanadium ions, as indeed observed in CaV$_2$O$_4$.

{\em Acknowledgement}. We acknowledge fruitful discussions with A.
Chubukov, D. Johnston, G. Japaridze,  O. Kolezhuk, B. Lake, O.
Pieper and O. Tchernyshyov.

\end{document}